\documentclass[aps,twocolumn,pre,showpacs,groupedaddress,amsmath,amssymb,superscriptaddress]{revtex4}
\usepackage{amsmath}
\usepackage{amssymb}
\usepackage{graphicx}
\usepackage{bm}
\usepackage{epic}
\usepackage{eepic}
\usepackage{pifont}
\usepackage[latin1]{inputenc}
\usepackage{rotating}
\usepackage{color}
\usepackage{nicefrac}
\usepackage{ulem}
\usepackage[caption=false]{subfig}

\input epsf
\input rotate
\usepackage{graphicx}

\newcommand{\rin}{R_{\text{in}}}
\newcommand{\rout}{R_{\text{out}}}
\newcommand{\Canum}{\text{Ca}}
\newcommand{\Renum}{\text{Re}}

\begin{document}
\draft
\title{Complex Dynamics of a Bilamellar Vesicle as a Simple Model for Leukocytes}
\author{Badr Kaoui}
\email{b.kaoui@tue.nl}
\affiliation{Department of Applied Physics, Eindhoven University of Technology, P. O. Box 513, 5600 MB Eindhoven, The Netherlands}
\author{Timm Kr\"{u}ger}
\affiliation{Centre for Computational Science, University College London, 20 Gordon Street, WC1H 0AJ London, United Kingdom}
\affiliation{Department of Applied Physics, Eindhoven University of Technology, P. O. Box 513, 5600 MB Eindhoven, The Netherlands}
\author{Jens Harting}
\affiliation{Department of Applied Physics, Eindhoven University of Technology, P. O. Box 513, 5600 MB Eindhoven, The Netherlands}
\affiliation{Institute for Computational Physics, University of
Stuttgart, Allmandring 3, 70569 Stuttgart, Germany}

\begin{abstract}
The influence of the internal structure of a biological cell (e.g., a
leukocyte) on its dynamics and rheology is not yet fully understood. By using 2D numerical simulations of a
bilamellar vesicle (BLV) consisting of two vesicles as a cell model, we find
that increasing the size of the inner vesicle (mimicking the nucleus) triggers
a tank-treading-to-tumbling transition. A new dynamical state is observed, the
\textit{undulating} motion: the BLV inclination with respect to the imposed
flow oscillates while the outer vesicle develops rotating lobes. The BLV
exhibits a non-Newtonian behavior with a time-dependant apparent viscosity
during its unsteady motion. Depending on its inclination and on its inner
vesicle dynamical state, the BLV behaves like a solid or a liquid.
\end{abstract}

\maketitle

\section{Introduction} 
\begin{figure*}
\centering
\subfloat[$0.25$]{\label{fig:figure01a}\includegraphics[angle=-90,width=0.1\textwidth]{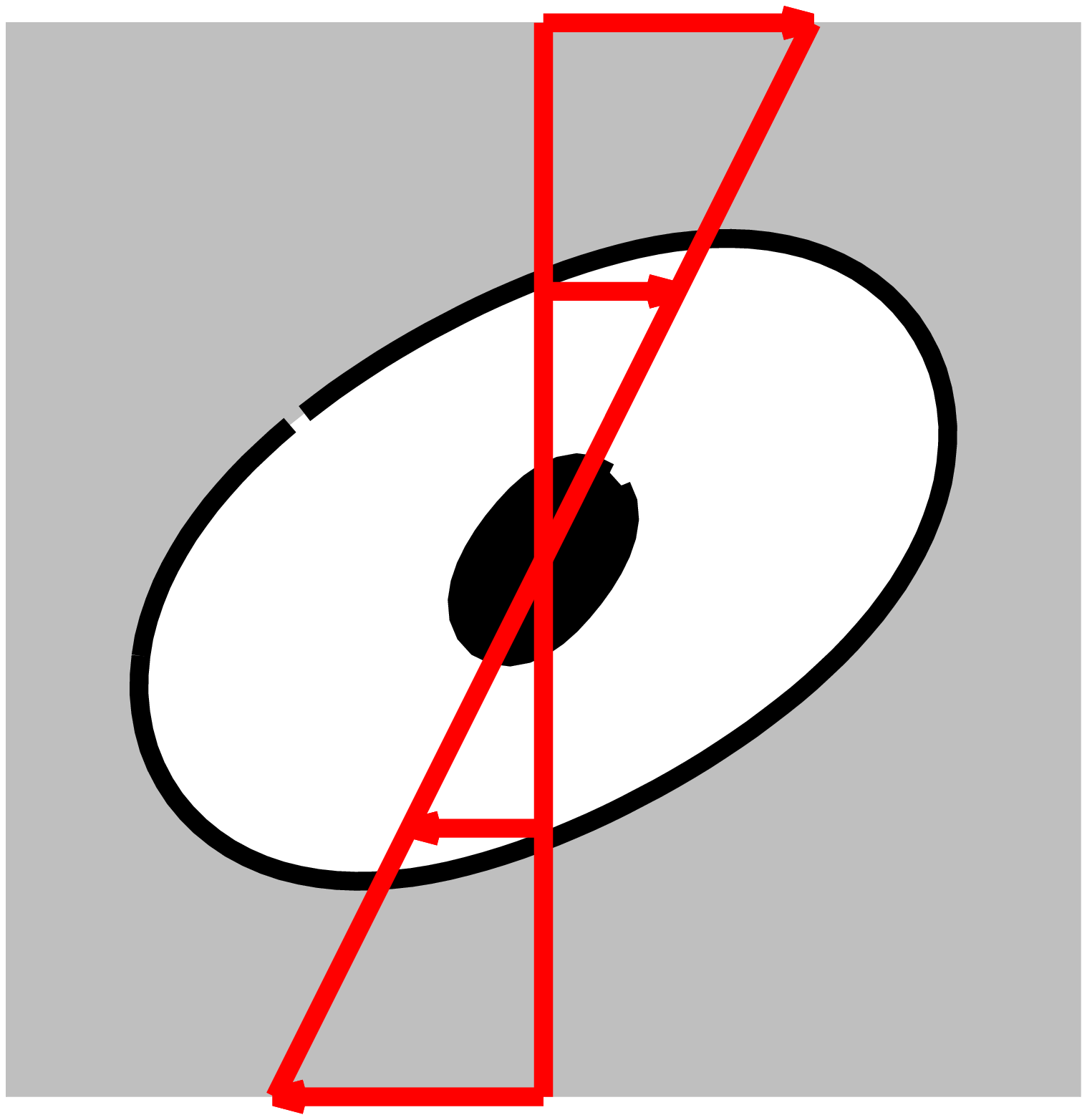}}
\quad
\subfloat[$0.40$]{\label{fig:figure01b}\includegraphics[angle=-90,width=0.1\textwidth]{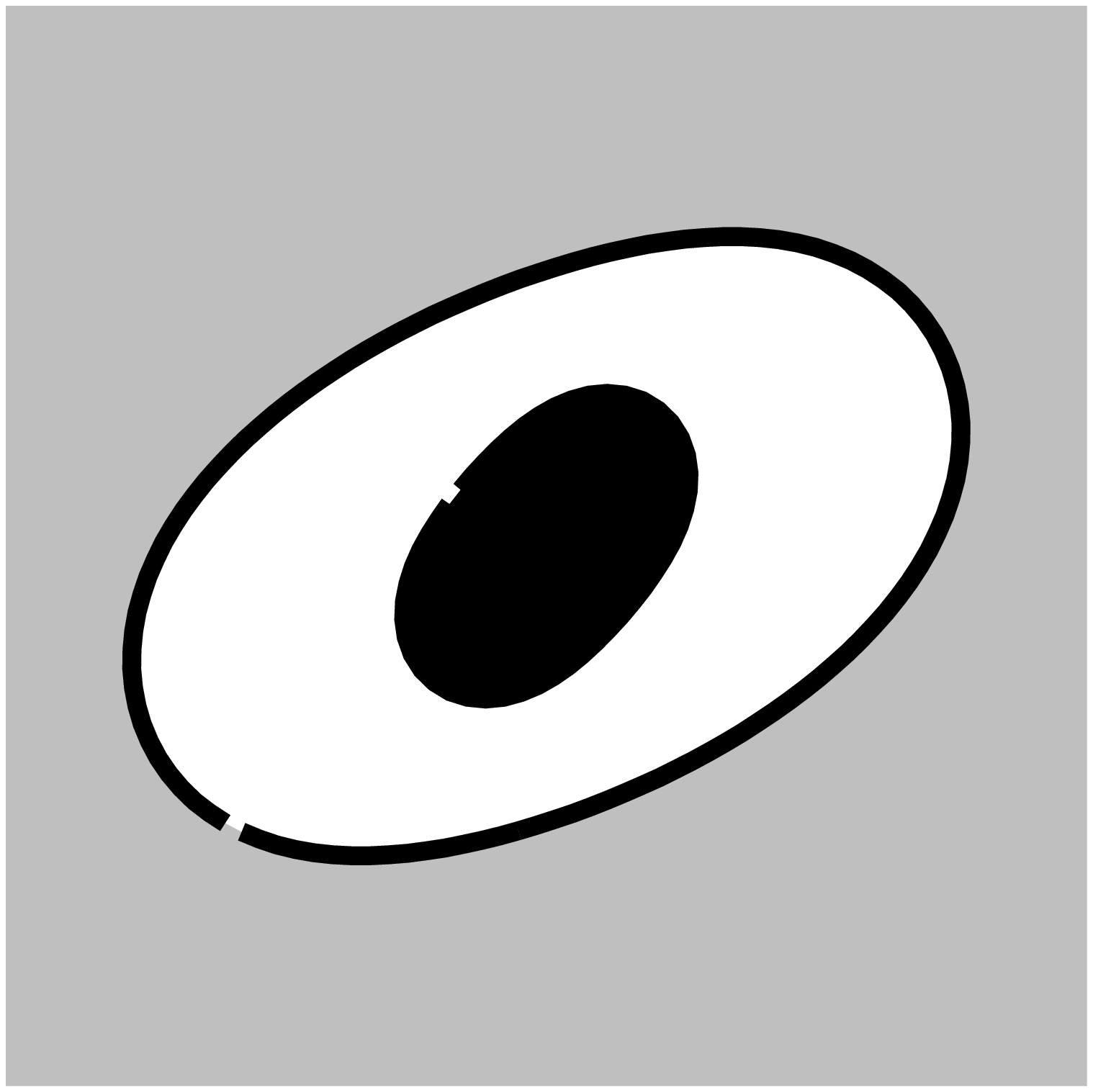}}
\quad
\subfloat[$0.55$]{\label{fig:figure01c}\includegraphics[angle=-90,width=0.1\textwidth]{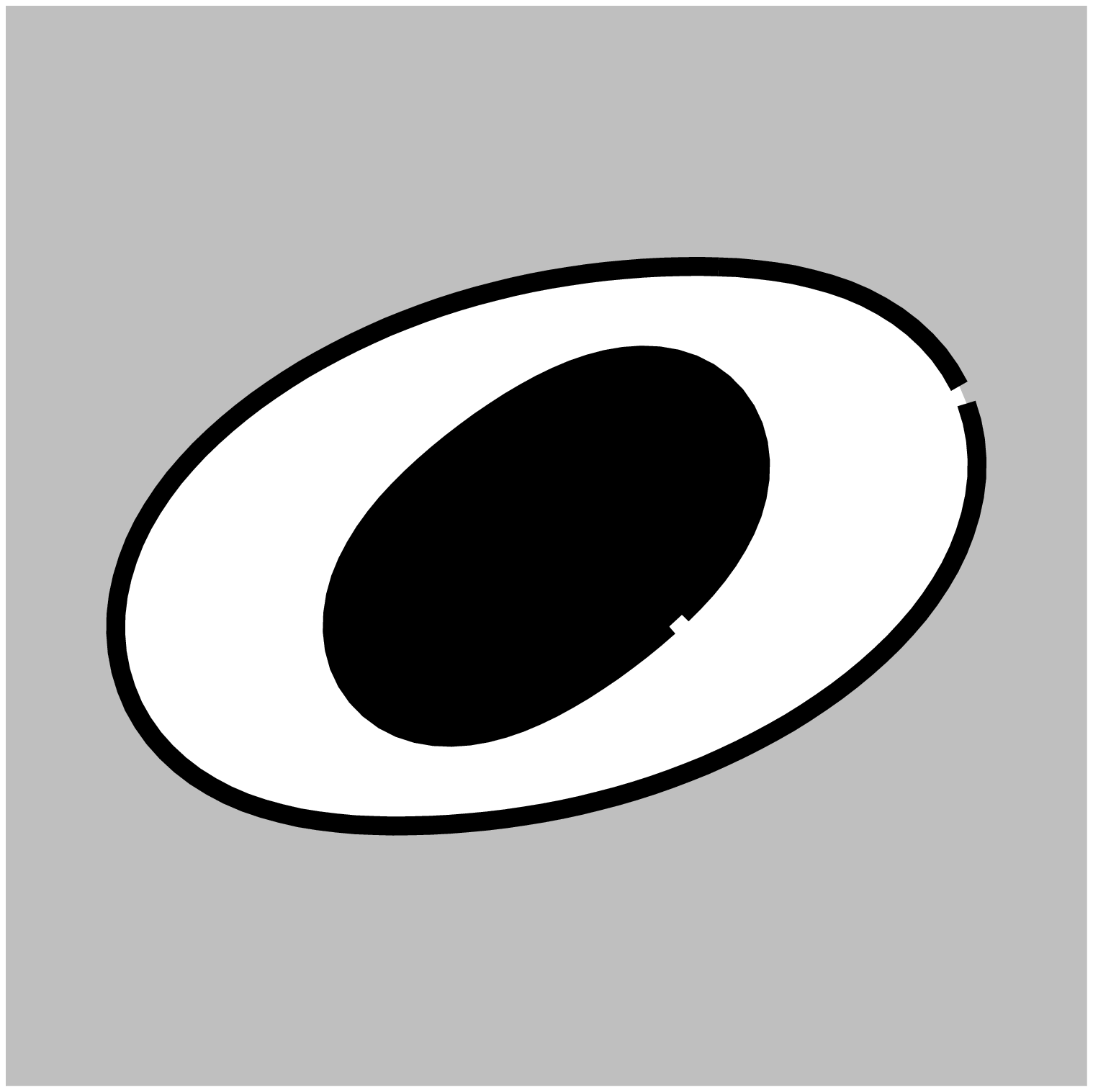}}
\quad
\subfloat[$0.75$]{\label{fig:figure01da}\includegraphics[angle=-90,width=0.1\textwidth]{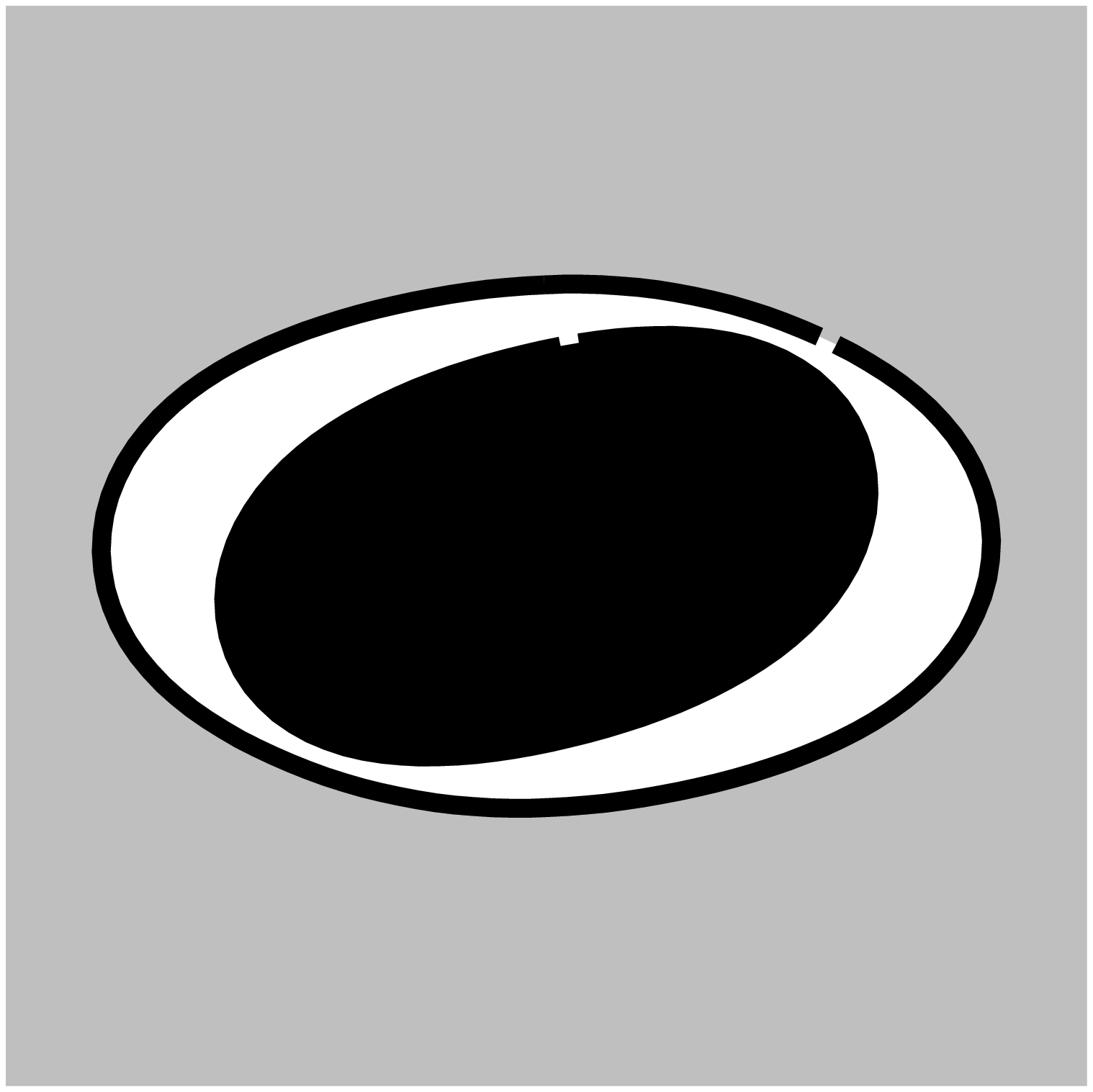}}
\subfloat{\label{fig:figure01db} \includegraphics[angle=-90,width=0.1\textwidth]{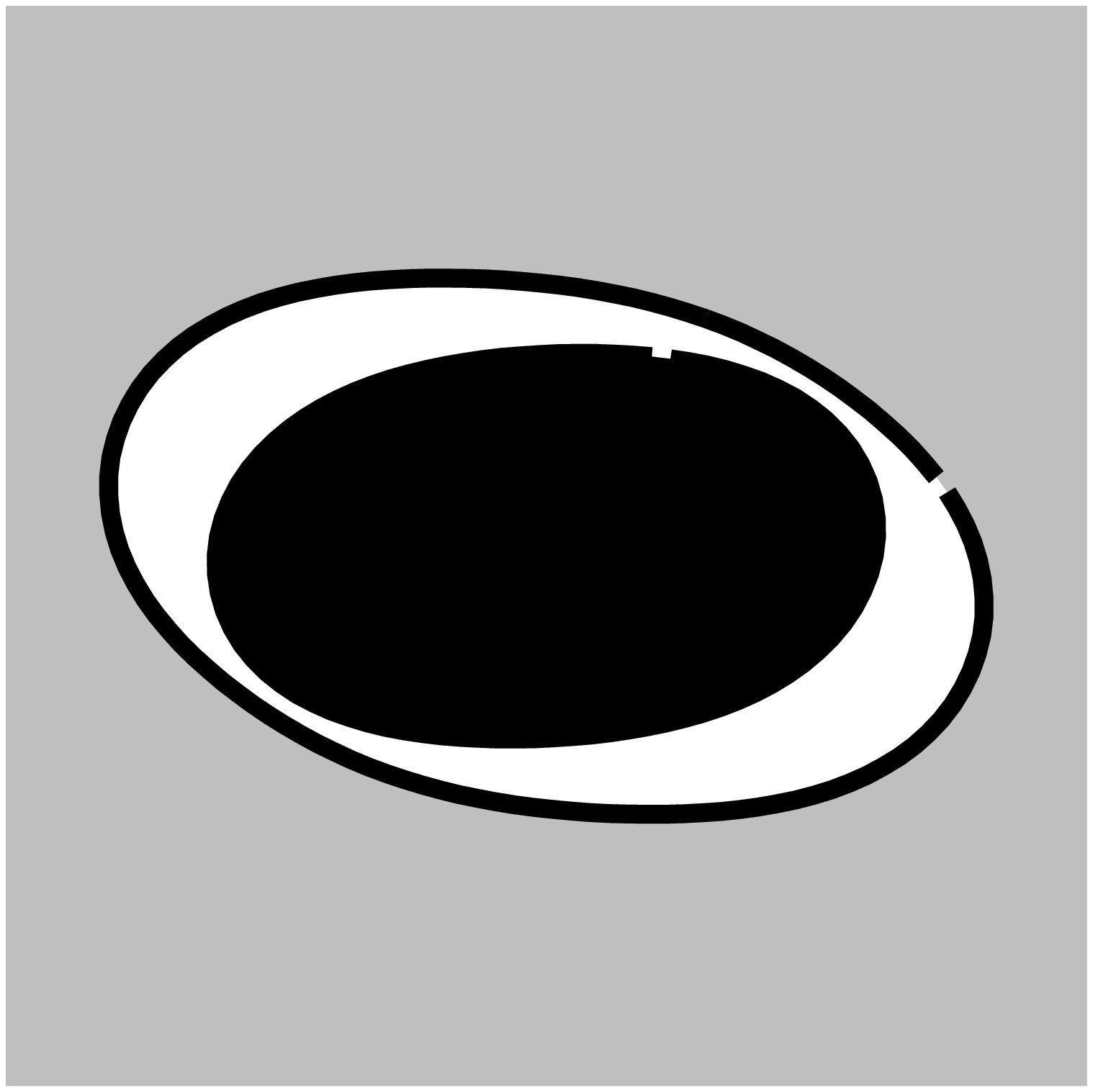}
\includegraphics[angle=-90,width=0.1\textwidth]{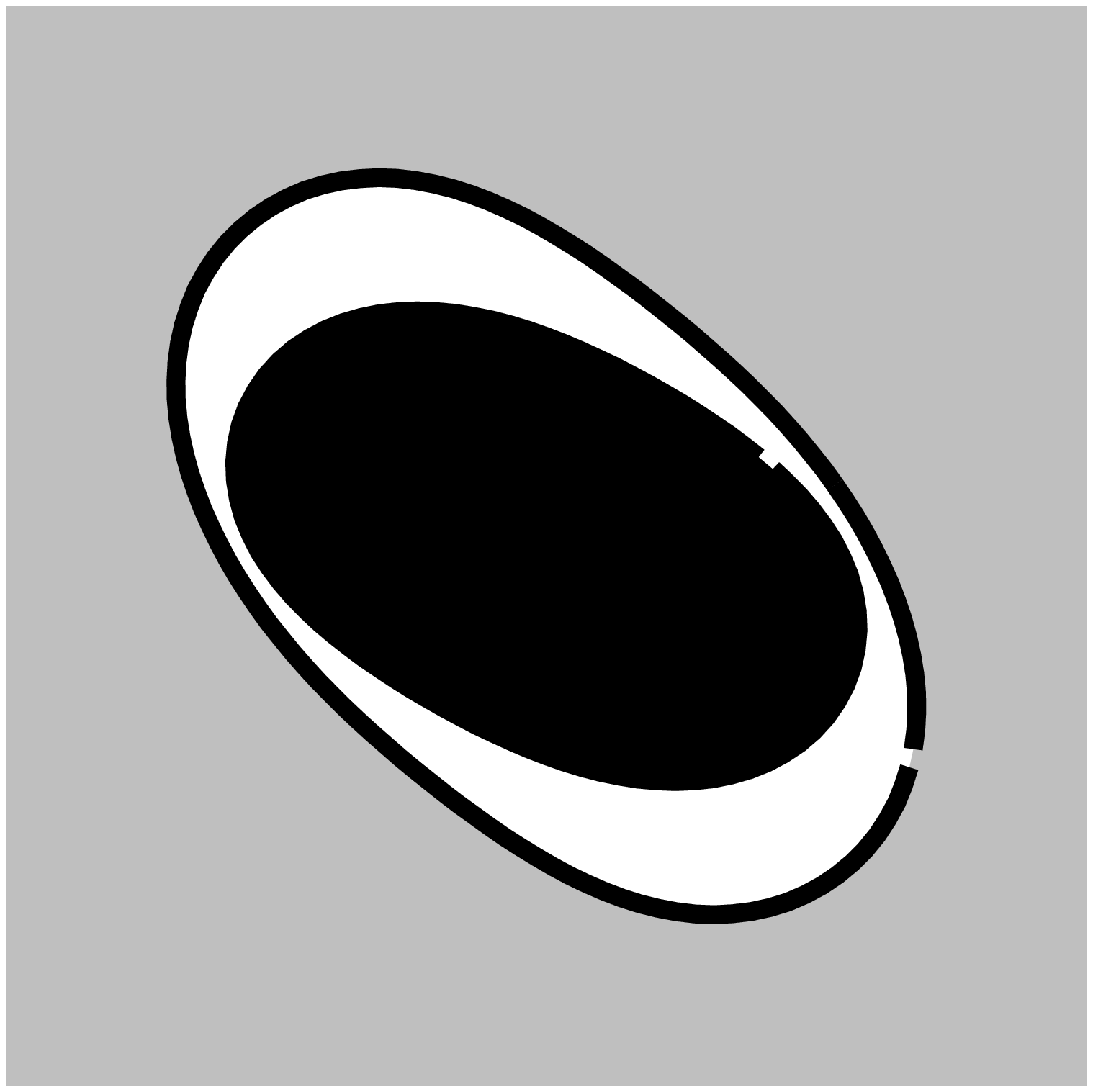}
\includegraphics[angle=-90,width=0.1\textwidth]{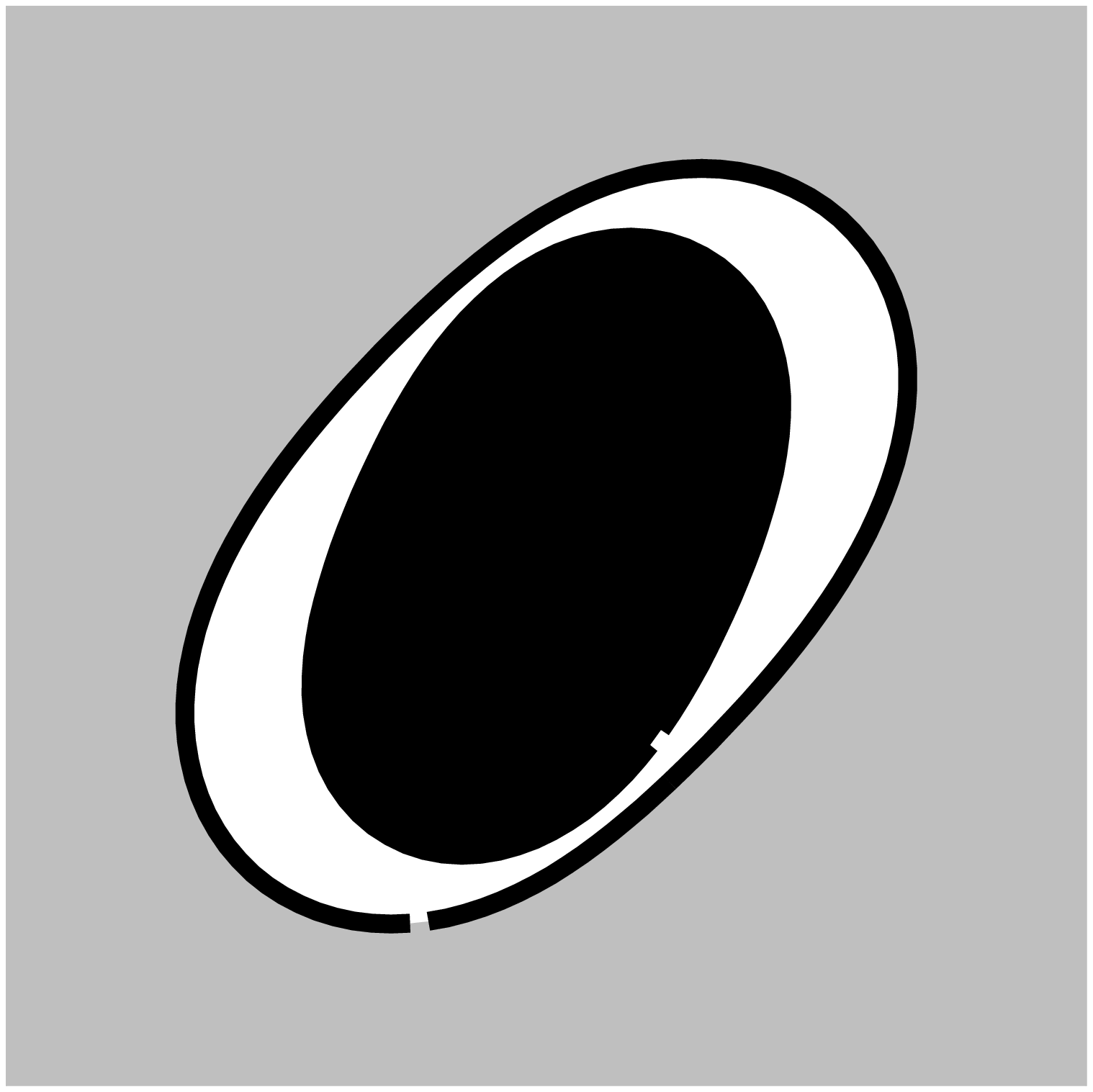}
\includegraphics[angle=-90,width=0.1\textwidth]{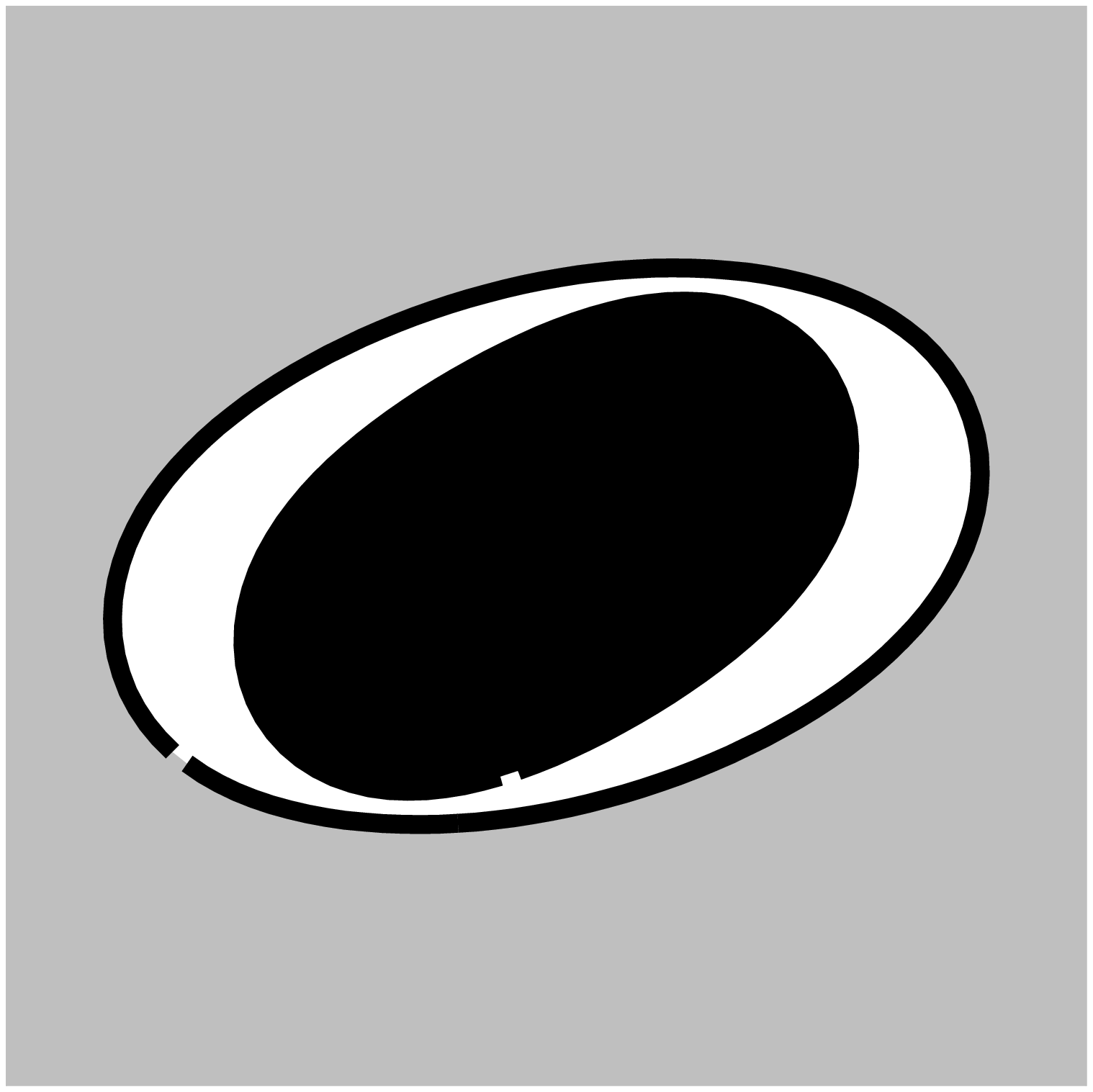}
\includegraphics[angle=-90,width=0.1\textwidth]{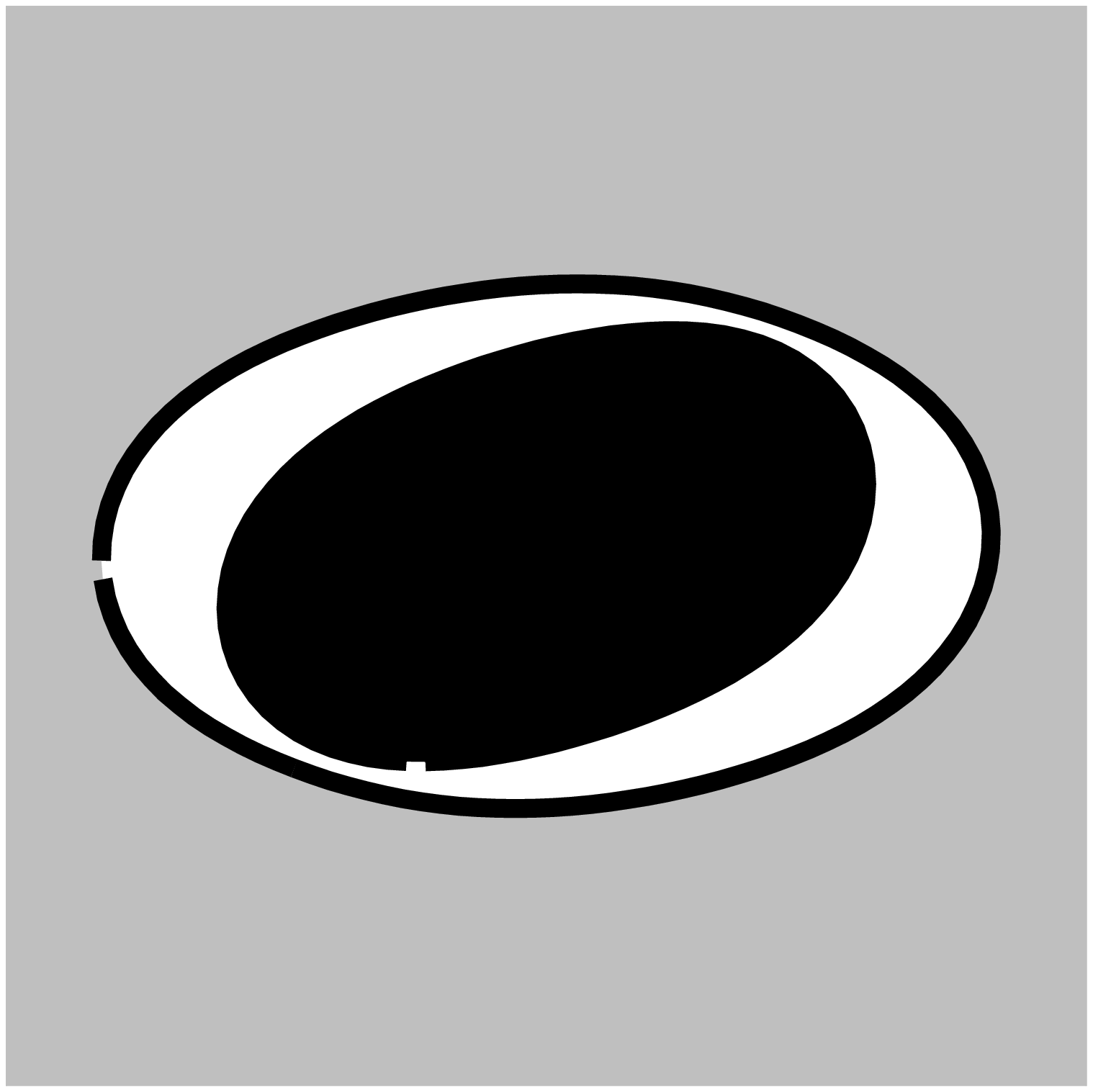}
}
\caption{\label{fig:figure01} Snapshots showing the dynamics of a
bilamellar vesicle (BLV) in shear flow in the low deformation regime
($\Canum = 0.5$). The ratio $\rin/\rout$ is given below each
image. For small inner vesicles (a)--(c), the BLV performs steady
tank-treading. The angle between the main axis of the BLV 
and the flow direction $\theta_\text{out}$ decreases with $\rin$ (see
Fig.~\ref{fig:figure02}). Beyond a threshold of $\rin$, the BLV starts
tumbling (d). This is similar to the tank-treading-to-tumbling transition
observed for viscous vesicles~\cite{Kantsler2006,Kaoui2012}, but is
triggered here (in the absence of a viscosity contrast) by the presence
of the inner vesicle.}
\end{figure*}

Unilamellar vesicles (ULVs) consisting of a
single closed phospholipid membrane were extensively used as biomimetic model
for erythrocytes (red blood cells) in the past. They succeeded to reproduce
many known features, like the steady shapes in Poiseuille
flow~\cite{Skalak1969,Kaoui2009} or the dynamical states under shear
flow~\cite{Fischer1978,Kraus1996,Kantsler2006}. However, for leukocytes (white
blood cells), despite their relevant role in the immune system, the dynamics
and rheology are still poorly understood since their
complex internal
structure dominated by the nucleus alters the mechanical properties in a non
straightforward manner~\cite{Tran-Son-Tay2007}.
We use a bilamellar vesicle (BLV) as a model for biological
cells, in particular, a leukocyte. While a ULV consists of a single vesicle, 
a BLV consists of two vesicles: an outer larger one (the cell) enclosing 
an inner smaller one (mimicking the nucleus)~\cite{Schmid-Schonbein1980}, see Fig.~\ref{fig:figure01}a. 
We study numerically the dynamics of a BLV under shear flow
and investigate how the dynamical and rheological properties of a leukocyte
are affected by varying the size and the deformability of
the nucleus as well as the amount of fluid enclosed between the nucleus and the
cell.
We show that leukocytes cannot be described simply by
fluid-filled particles enclosing a homogeneous fluid without an internal
structure as it has been used, for example, in
Refs.~\cite{Cantat1999,Fedosov2012}. This is because leukocytes adapt their
mechanical properties and act as a solid or as a liquid depending on
how they are deformed by the imposed fluid~\cite{Tran-Son-Tay2007}.

\section{Simulation method} 
We consider two concentric vesicles in 2D in a
shear flow generated between two parallel plates. We designate by $\rout$ and
$\rin < \rout$ the effective radii of the outer and the inner vesicle ($R = P /
(2\pi)$, where $P$ is the vesicle perimeter). All fluids are considered to be
incompressible, Newtonian and of the same viscosity $\eta$. 
Their flow is solved by the
lattice-Boltzmann method and the fluid-vesicle two-way coupling is
achieved employing the immersed boundary method (see~\cite{Kaoui2011a} for
details). Both vesicle membranes are locally inextensible and
experience resistance to bending with the same
rigidity $\kappa$. They exert a reaction force per unit length (in 2D)
\begin{equation}
\mathbf{f} \!=\! \left[\! \kappa \!\left(\frac{\partial^2 c}{\partial s^2} +
\frac{{c}^3}{2}\right) \!-\! c \sigma ^* \right]\!\mathbf{n} +
\frac{\partial\sigma ^*}{\partial s}\mathbf{t}
\label{eq:force}
\end{equation}
on the surrounding fluid where $\mathbf{n}$ and $\mathbf{t}$ are the
unit normal and tangent vectors, $c$ is the local curvature, $s$ is the
curvilinear coordinate and $\sigma^*$ is the local effective surface
tension. Both membranes interact purely hydrodynamically.
The distance between the plates is chosen as such
that the effect of wall confinement is negligible~\cite{Kaoui2011a,Kaoui2012}.

First, we investigate how the dynamics of the BLV is affected by varying two parameters: i) $\rin$, to study the effect of its internal structure and ii) the deformability number $\Canum = \eta \gamma \rout^3 / \kappa$, which we
define in the style of a capillary number used for droplets, but based on the
bending rigidity instead of the surface tension. Here, $\gamma$ is the shear
rate.
Second, we investigate how the
dynamics of the inner vesicle is affected by the flow induced by the outer
one by varying the swelling degree $\Delta_\text{out}=4
\pi A_\text{out} / P_\text{out}^2$ ($A_\text{out}$ is the outer vesicle
area) while keeping all other parameters fixed. All simulations are
performed in the Stokes regime:
$\mathcal{O}(\Renum) = 10^{-2}$, where $\Renum =\rho \gamma \rout^2 / \eta$ is
the Reynolds number and $\rho$ is the fluid density. Both vesicles are deflated
and have a swelling degree $\Delta_\text{in} = \Delta_\text{out} = 0.9$. While
this is a typical situation for vesicles, for leukocytes it corresponds to
large deformation encountered in capillaries or in micropipette
experiments~\cite{Tran-Son-Tay2007}. We vary $\rin$ while keeping $\rout$
fixed.
\begin{figure}[b]
\centerline{\includegraphics[width=.18\textwidth,angle=-90]{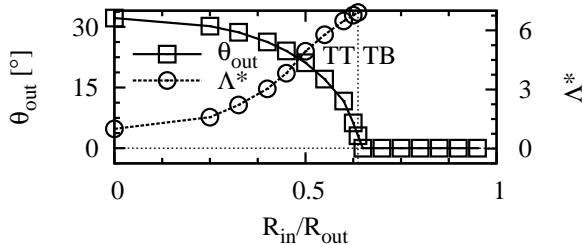}}
\caption{\label{fig:figure02}
The inclination angle $\theta_\text{out}$ (left axis) and the apparent
viscosity contrast $\Lambda ^*$ (right axis) as a function of
the radius ratio in the low deformation regime (${\rm Ca}=0.5$).
An increase of $R _{\rm in}/R _{\rm out}$
induces an increase in $\Lambda ^*$ and the BLV becomes more and more
viscous. Consequently, $\theta _{\rm out}$ decreases until it vanishes
at the transition point $R_{\rm int}/R_{\rm out}=0.64$.}  
\end{figure}
\begin{figure}[t]
\centerline{\includegraphics[width=.18\textwidth,angle=-90]{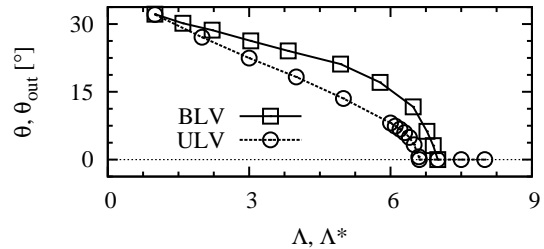}}
\caption{\label{fig:figure03}
The inclination angle \textit{vs.} the viscosity contrast for a
viscous ULV and a BLV in the low deformation regime (${\rm Ca}=0.5$): the angle decreases with increasing
$\Lambda$ or $\Lambda ^*$. However, for a given $\Lambda$ or $\Lambda ^*$,
the angle of the BLV is larger than that of the viscous ULV demonstrating
a difference in the quantitative behavior of both.}
\end{figure}

\section{Observations and discussion} 
The obtained dynamics is shown in Fig.~\ref{fig:figure01} as a function of
$0 < \rin / \rout < 1$.  For smaller inner vesicles
(Fig.~\ref{fig:figure01}(a)--(c)), both vesicles perform a steady
tank-treading motion (they assume a steady inclination angle with respect to the flow while their membranes undergo a tank-treading like motion).
The BLV aligns more and more with the flow when $\rin$ is
increased.  Due to the symmetry, both vesicle centers are stationary.
Beyond a threshold of $\rin$, the motion of the BLV transits from steady
tank-treading to unsteady tumbling motion (rotation as solid elongated particle).
Fig.~\ref{fig:figure01}(d) shows snapshots of a tumbling BLV with $\rin /
\rout = 0.75$. The inner vesicle assumes a relative angle with respect to
the main axis of the outer one.
During the tumbling motion, the mean value of
$\theta_\text{out}$ (the angle defined by the main long axis of the BLV
and the flow direction) is zero. When plotting $\theta_\text{out}$ as
function of $\rin / \rout$ (Fig.~\ref{fig:figure02}), we see that
$\theta_\text{out}$ decreases with increasing $\rin / \rout$ until it
vanishes at a critical value where tank-treading (TT) is
replaced by tumbling (TB). The TT-TB transition is known
for viscous ULVs where it is induced by increasing the viscosity
contrast $\Lambda$ (ratio between the internal and external fluid
viscosities) beyond a given threshold $\Lambda _\text{cr}$~\cite{Kantsler2006,Kaoui2012}.
Here, however, all fluids have the same viscosity and the transition is
induced solely by the presence of the encapsulated vesicle and by enlarging its size.
Veerapaneni \textit{et al.}~\cite{Veerapaneni2011} predicted a similar transition for non-viscous ($\Lambda=1$) ULVs
enclosing a solid particle.
They claim that the inclusion increases the \textit{apparent} internal
viscosity leading to the transition as observed for inclusion-free ULVs
with $\Lambda _\text{cr} > 1$. To investigate the effect of the apparent internal viscosity, we follow \cite{Batchelor1970,Bagchi2010} and compute $\eta^* = \eta + \langle \sigma_{xy}\rangle / \langle S_{xy}\rangle$, where $\langle \sigma _{xy} \rangle = -\oint_{\partial \Omega_\text{in}} \text{d}s\, (f_x r_y) / A_\text{in}$ is the average excess shear stress caused by the presence of the inner vesicle. Therefore, the integration has to be performed on the surface of $\Omega_\text{in}$. The average shear rate within the outer vesicle domain $\Omega_\text{out}$ (consisting of the region between the two vesicles and the region within the inner vesicle) can be written as a surface integral by making use of Gauss's theorem: $\langle S_{xy} \rangle = \oint_{\partial \Omega_\text{out}} \text{d}s\, (n_x u_y + n_y u_x) / A_\text{out}$.
Here, $\mathbf r$ and $\mathbf u$ are the position and the velocity of a membrane element, respectively.
The apparent viscosity contrast of the BLV, $\Lambda^* = \eta^* / \eta$,
{\it vs.} $\rin / \rout$ is depicted
in Fig.~\ref{fig:figure02}: increasing $\rin$ leads to a monotonic increase
of $\Lambda^*$. The TT-TB transition for the BLV takes place at a
critical value of $\Lambda^*_\text{cr} = 6.9$. This value is close to the
critical viscosity contrast $\Lambda_\text{cr} = 6.6$ required for a
viscous ULV to undergo the same transition (for the same swelling degree
of $0.9$). A systematic comparison of the inclination angle ($\theta$,
$\theta_\text{out}$) \textit{vs.} the viscosity contrast ($\Lambda$,
$\Lambda^*$) of a viscous ULV and a BLV (see Fig.~\ref{fig:figure03})
shows that both exhibit similar qualitative behavior: the angle
decreases with increasing viscosity contrast until it vanishes at the
transition point. However, for all viscosity contrasts, the
angle of the BLV is found to be larger than that of the viscous ULV
($\theta_\text{out} > \theta$), especially at larger $\Lambda^*$
corresponding to larger inner vesicles. This demonstrates that a BLV
does not behave exactly as a viscous ULV for which the internal fluid is a
homogeneous medium. An internal heterogeneous medium, as it is the case for a leukocyte, with viscosity contrasts between the intranucleus fluid, the cytoplasm and the plasma, would affect the critical value of the dynamical transition. For example, a tank-treading BLV with a given ratio $R_{\rm in}/R_{\rm out}$ is expected to transit to tumbling only by making the inner vesicle fluid more viscous.

\begin{figure}[h]
\centerline{\includegraphics[width=0.3\textwidth,angle=270]{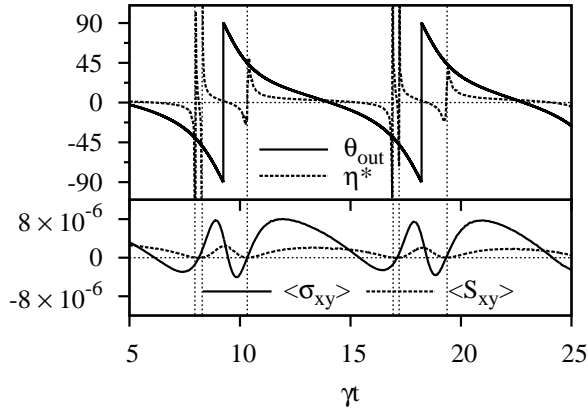}}
\caption{\label{fig:figure04}
Time evolution of the inclination angle $\theta_\text{out}$ (in degrees)
and the apparent internal viscosity $\eta^*$ (in lattice units) of a tumbling BLV 
($\rin / \rout = 0.75$, $\Canum = 0.5$). The data corresponds 
to the snapshots in Fig.~1(d). The time evolution of the shear stress
$\langle \sigma_{xy} \rangle$ and shear rate $\langle S_{xy} \rangle$ is
also shown (in lattice units). The time dependence of $\eta^*$ suggests non-Newtonian fluid properties of the inner medium (fluid and inner vesicle). Note especially $\eta^* \rightarrow
\infty$ (solid limit) when $\theta = \pm \pi /4$ (i.e. $\langle S_{xy}
\rangle \rightarrow 0$).}
\end{figure}

Moreover, when tumbling, the
apparent internal viscosity $\eta ^*$ of the BLV is a time-dependent quantity and varies in a coherent
manner with $\theta_\text{out}$, see Fig.~\ref{fig:figure04}. It diverges to the limit of a solid medium when
$\theta_\text{out} = \pm \pi / 4$ (direction of the
elongation/compression of the shear flow) since the average shear rate
$\langle S_{xy} \rangle$ vanishes at that point. This is a signature of
the non-Newtonian rheological behavior.
The BLV internal medium changes its apparent viscosity as
a response to the orientation with respect to the
flow. For biological systems, e.g., a leukocyte flowing in a vessel (Poiseuille
flow), this behavior suggests that the apparent viscosity depends on the
the stresses experienced by the cell, which vary with the instantaneous
lateral position within the vessel. 
\begin{figure}[t]
\centerline{\includegraphics[width=0.26\textwidth,angle=270]{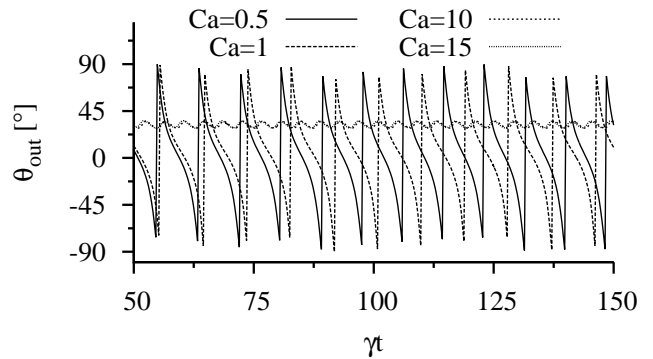}}
\caption{\label{fig:figure05}
The time evolution of the outer inclination angle $\theta_{\rm out}$ of the BLV for different values of ${\rm Ca}$. The inner vesicle radius is sufficiently large ($R_{\rm in}/R_{\rm out}=0.83$) and both vesicles have the same swelling degree $\Delta_{\rm in}=\Delta_{\rm out}=0.9$. For the same structural parameters, only by varying the degree of deformation ${\rm Ca}$ from $1$ to $15$, the BLV ceases to tumble and transits to the undulating state.}
\end{figure}
\begin{figure}[b]
\centering
\subfloat[]{\includegraphics[angle=-90,width=0.44\textwidth]{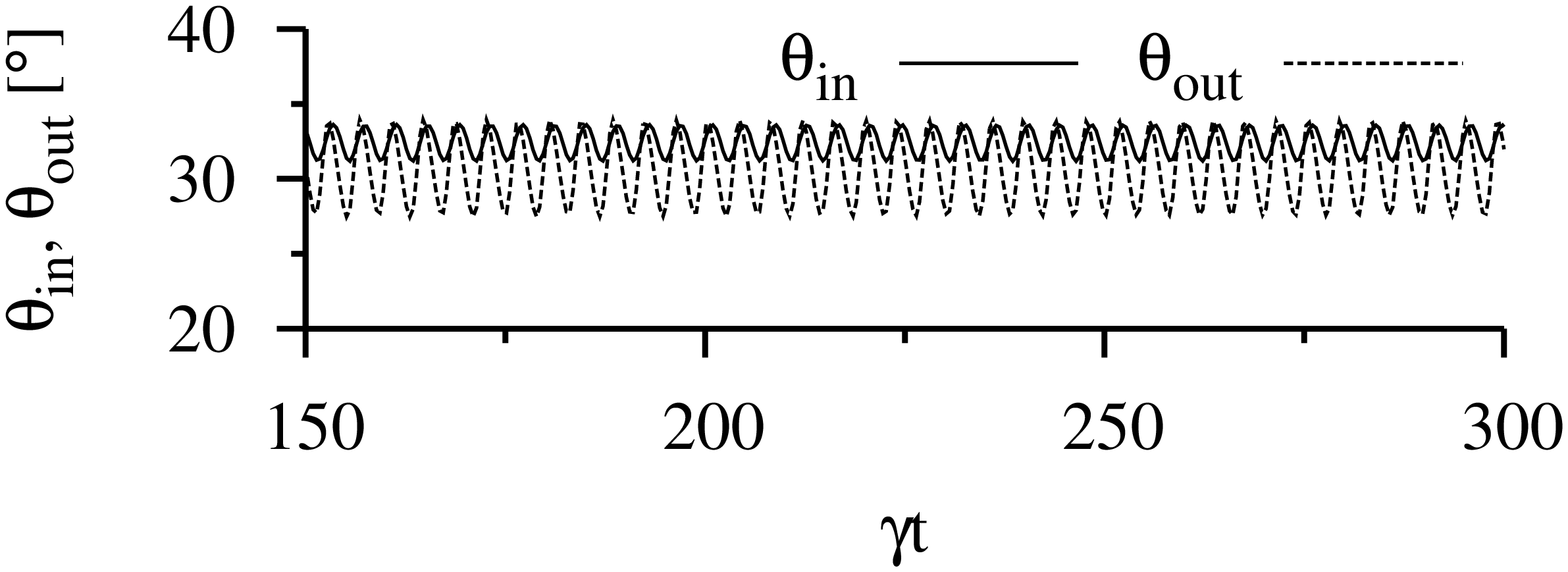}}
\\
\subfloat[]{\includegraphics[angle=-90,width=0.088\textwidth]{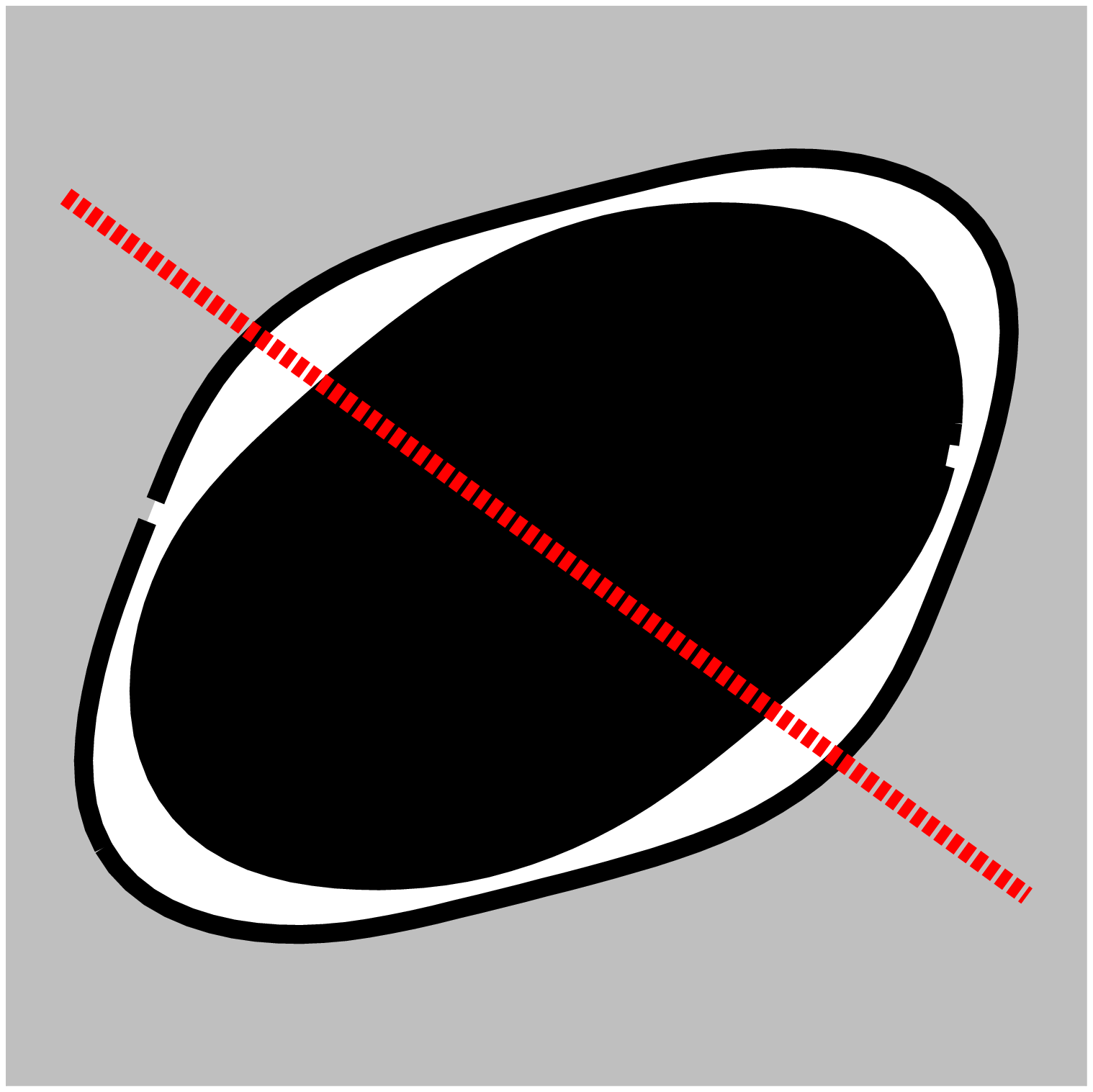}
\includegraphics[angle=-90,width=0.088\textwidth]{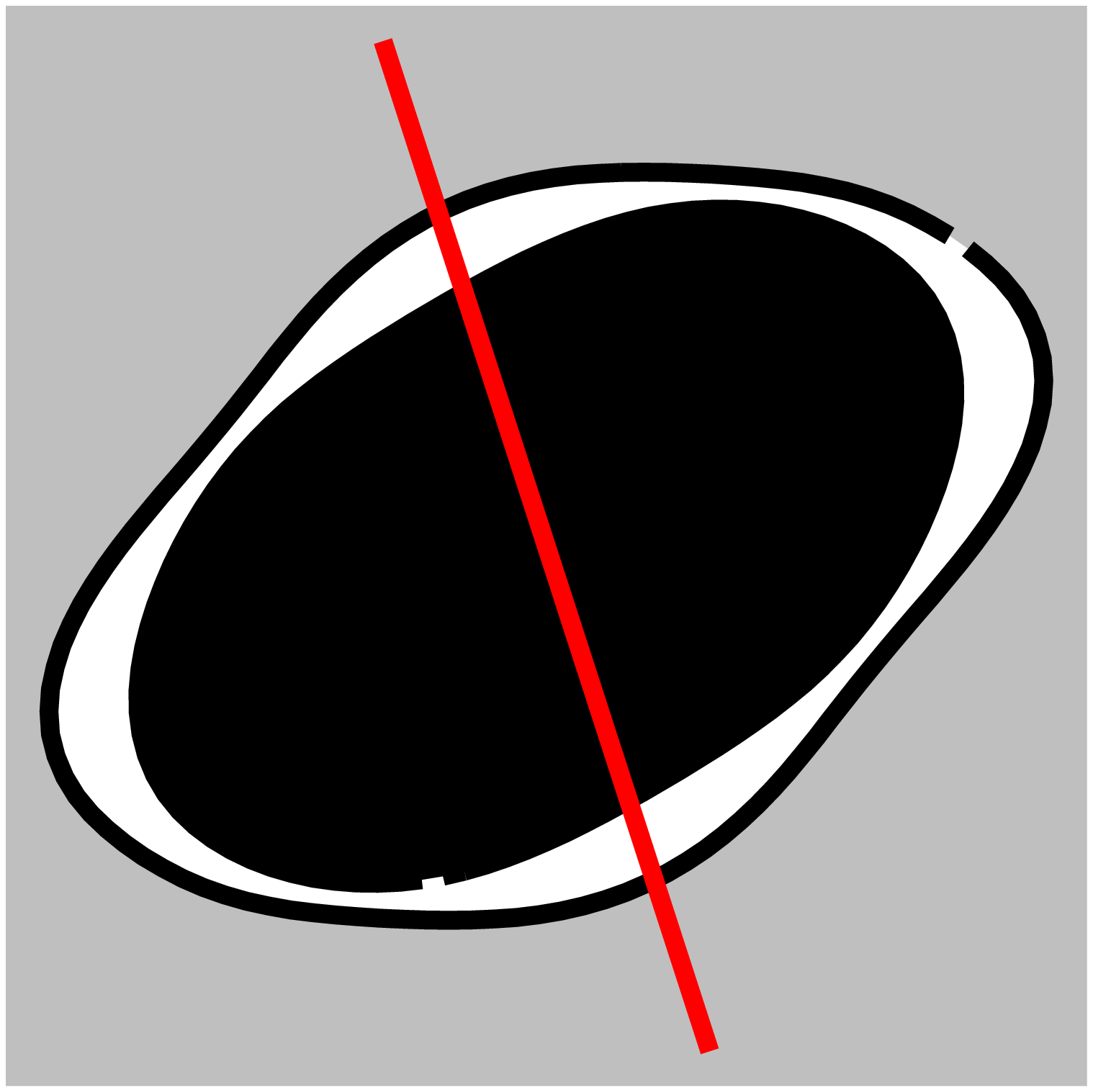}
\includegraphics[angle=-90,width=0.088\textwidth]{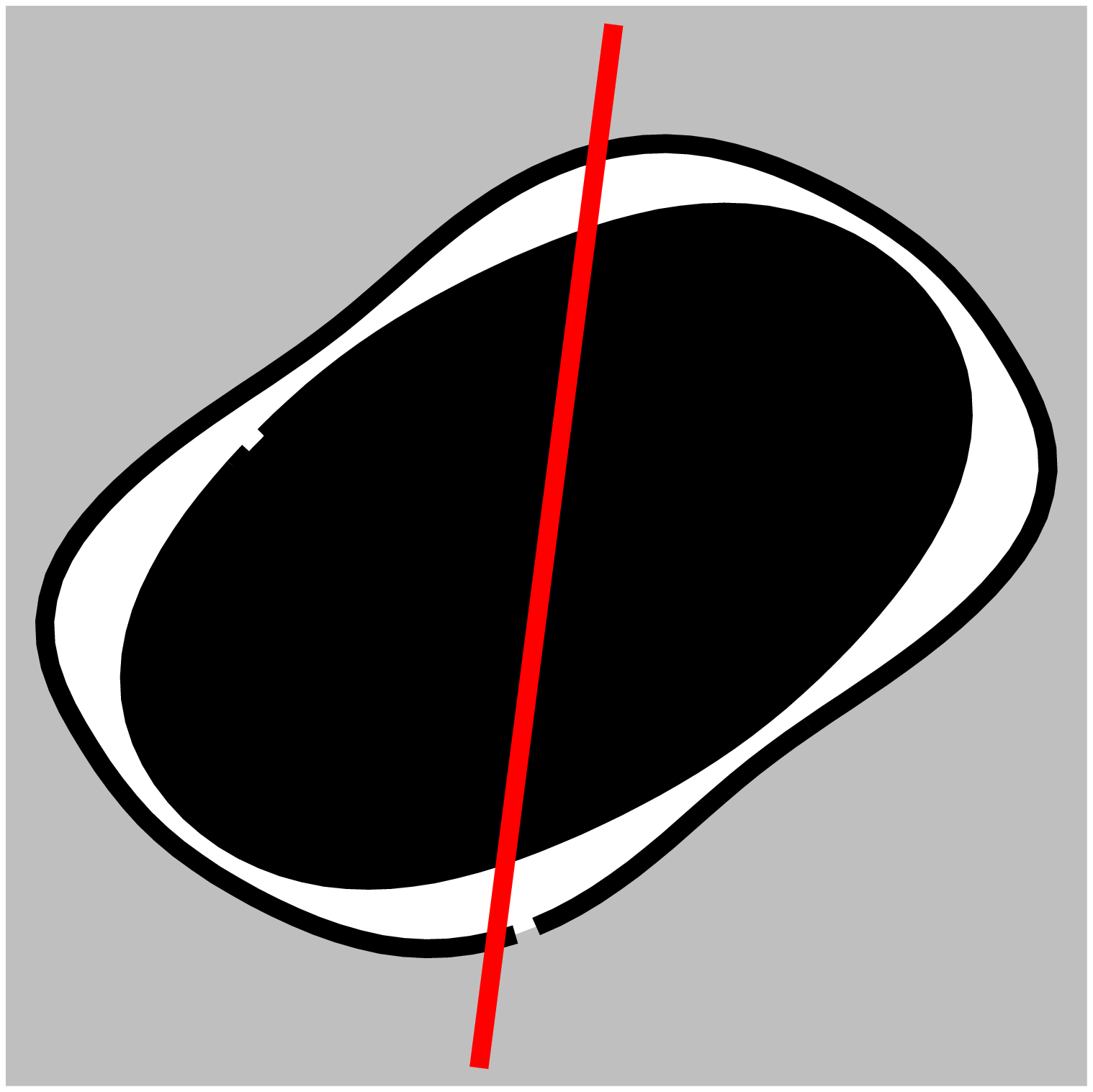}
\includegraphics[angle=-90,width=0.088\textwidth]{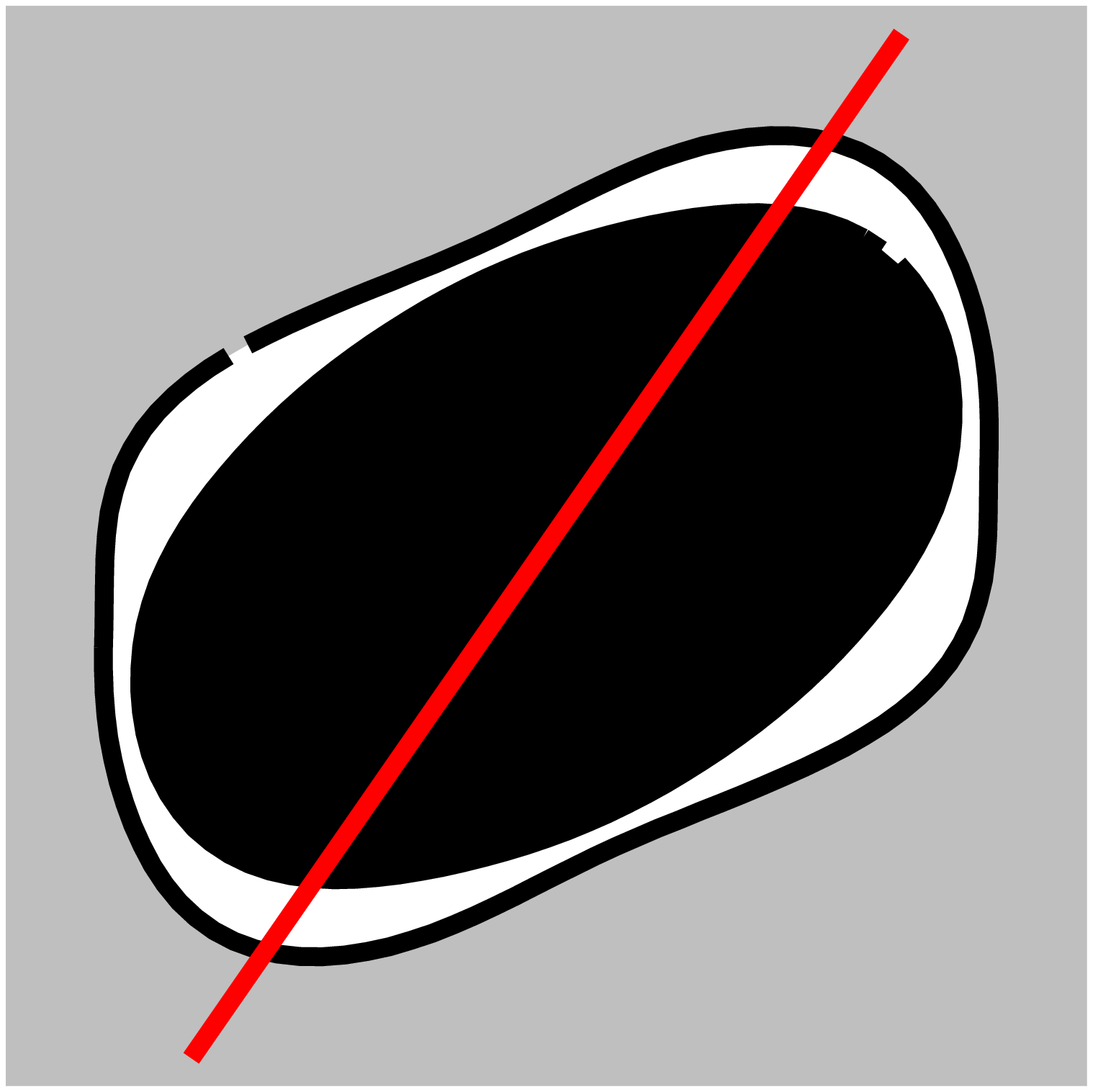}
\includegraphics[angle=-90,width=0.088\textwidth]{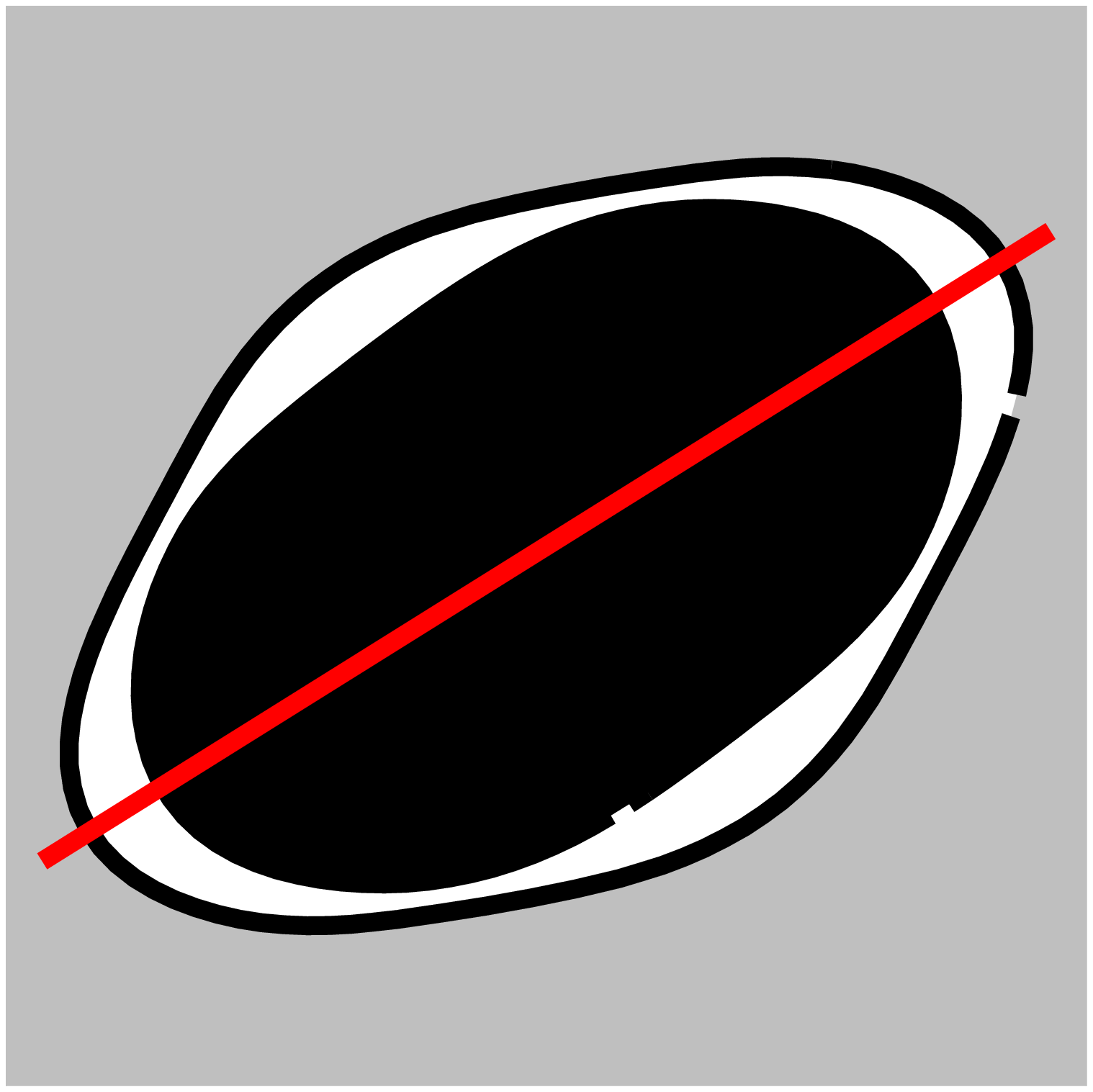}}
\caption{\label{fig:figure06} (a) Oscillations of the main axes of the
inner ($\theta _{in}$) and outer ($\theta _{out}$) vesicles of a BLV
during the undulating motion ($\rin/ \rout = 0.85$, $\Canum = 10$). 
(b) Snapshots, taken
at equal time intervals, showing four lobes of the outer vesicle membrane
and their rotation. The straight line
denotes the consequent locations of two opposite lobes. The undulating regime
replaces tumbling for larger $R _{\rm in}/R _{\rm out}$ and larger ${\rm Ca}$.}
\end{figure}

At higher $\Canum$, the membrane deformability becomes important and thus the
BLV deforms substantially. For a smaller inner vesicle, the BLV again
tank-treads. However, for larger $\rin$, in contrast to the limit
of small deformation, we surprisingly observe that the
BLV does not tumble anymore. It rather performs a new type of unsteady
motion (Fig.~\ref{fig:figure05}). The inner vesicle undergoes a
\textit{swinging} motion; the main axis oscillates about a positive mean
angle ($\theta _{\rm in}$ in Fig.~\ref{fig:figure06}(a)) while its shape
does not deform.  Such motion is known for red blood cells and capsules~\cite{Abkarian2007}.
The outer vesicle exhibits a non-regular motion: although its main long
axis performs oscillations about a positive mean angle as well
($\theta _{\rm out}$ in Fig.~\ref{fig:figure06}(a)), its shape undergoes
larger undulations, i.e., its membrane buckles. It
develops two oscillating lobes for intermediate sized inner vesicles or
four rotating lobes for larger inner vesicles
(Fig.~\ref{fig:figure06}(b)). Both unsteady motions cannot be qualified
as \textit{vacillating-breathing} (VB)~\cite{Misbah2006},
\textit{swinging} (SW) or \textit{trembling} (TR)~\cite{Bagchi2012}. An
almost similar feature as shown in
Fig.~\ref{fig:figure06} has been recently observed experimentally by
Pommella \textit{et al.} \cite{Pommella2012} for a surfactant
multilamellar droplet subjected to strong shear. The authors
describe the droplet motion as VB, but we disagree with this
classification since the angle is found to oscillate about a non-zero mean
value as it does for SW.
Yet, SW can be ruled out as well since it is observed in the small
deformation limit.
The dynamic mode of the BLV can neither be described as TR which
indeed is characterized by the formation of lobes \cite{Bagchi2012} but also
requires the shape to become perfectly elliptical at a certain moment.
This is impossible for BLV due to the presence of the large inner obstacle.
The appearance of this new unsteady motion (for larger $\Canum$ and larger $\rin /
\rout$) that we name \textit{undulating} motion cannot be explained solely
based on the apparent viscosity contrast argument. The inner vesicle disturbs
the motion of the outer one. By increasing its size, the thickness of the
fluid layer between the membranes decreases to become a thin liquid film.
The outer membrane tries to tank-tread under the external applied shear.
However, the presence of the inner vesicle prevents this and thus it slides
over the inner membrane, which plays the role of a nearly solid obstacle. Its
shape is less deformable because i) the inner vesicle is smaller ($\Canum
\propto R^3$), and ii) the outer vesicle shields the inner one from the external
flow.  A thorough understanding of the appearance of the undulating motion is
still missing. However, a relation to the Marangoni effect 
can be proposed: at large deformations, bending becomes less important than
tension.  We observe a non-uniform distribution of the surface tension for the
BLV, $\partial \sigma^* / \partial s \neq 0$.
This is in line with the surfactant multilamellar droplet \cite{Pommella2012} and the instability of thin liquid films on a solid substrate \cite{Pozrikidis2011}.
\begin{figure}[t]
\centering
\subfloat[$\Delta_\text{out} = 0.90$]{\label{fig:figure07a}\includegraphics[angle=-90,width=0.14\textwidth]{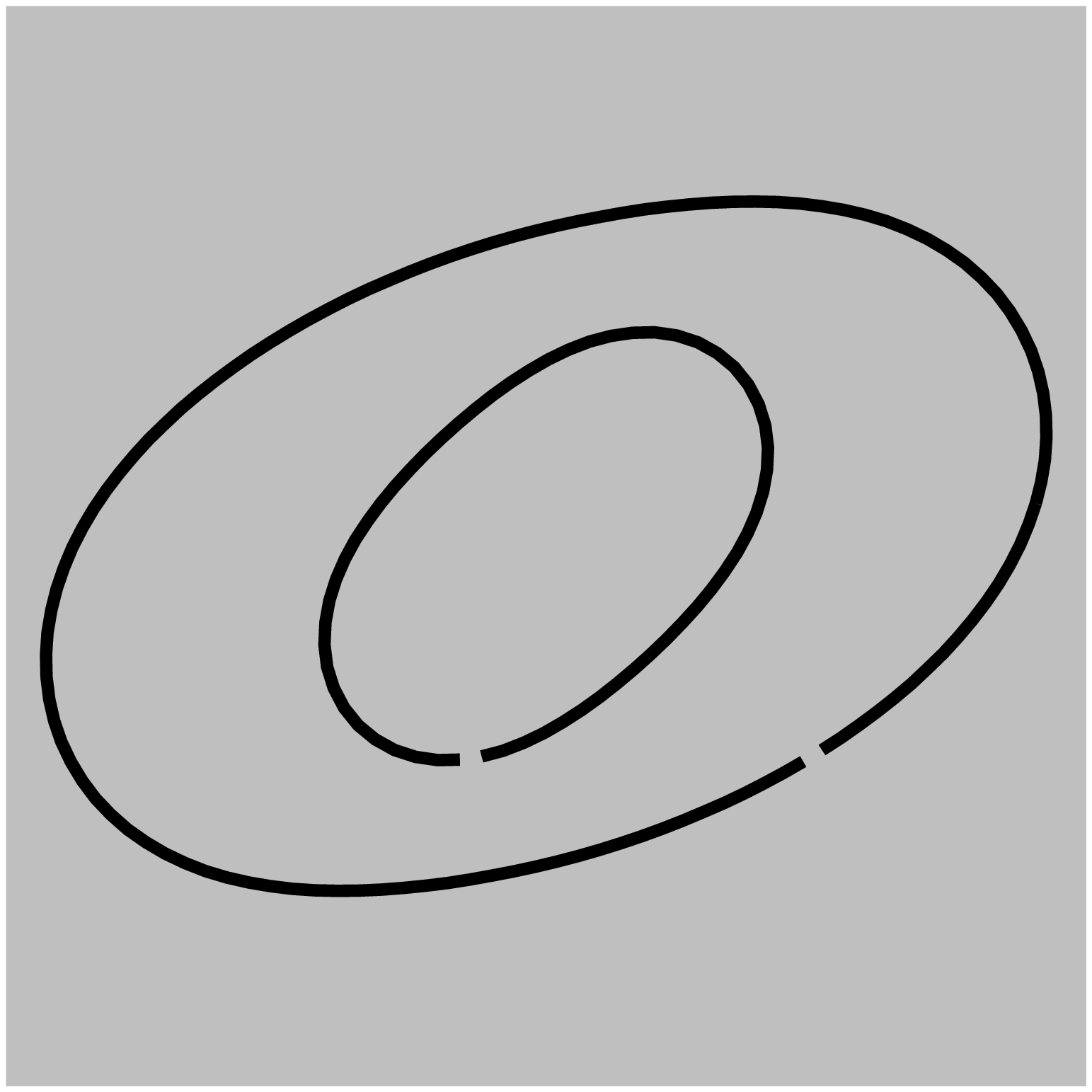}}
\quad
\subfloat[$\Delta_\text{out} = 0.98$]{\label{fig:figure07b}\includegraphics[angle=-90,width=0.14\textwidth]{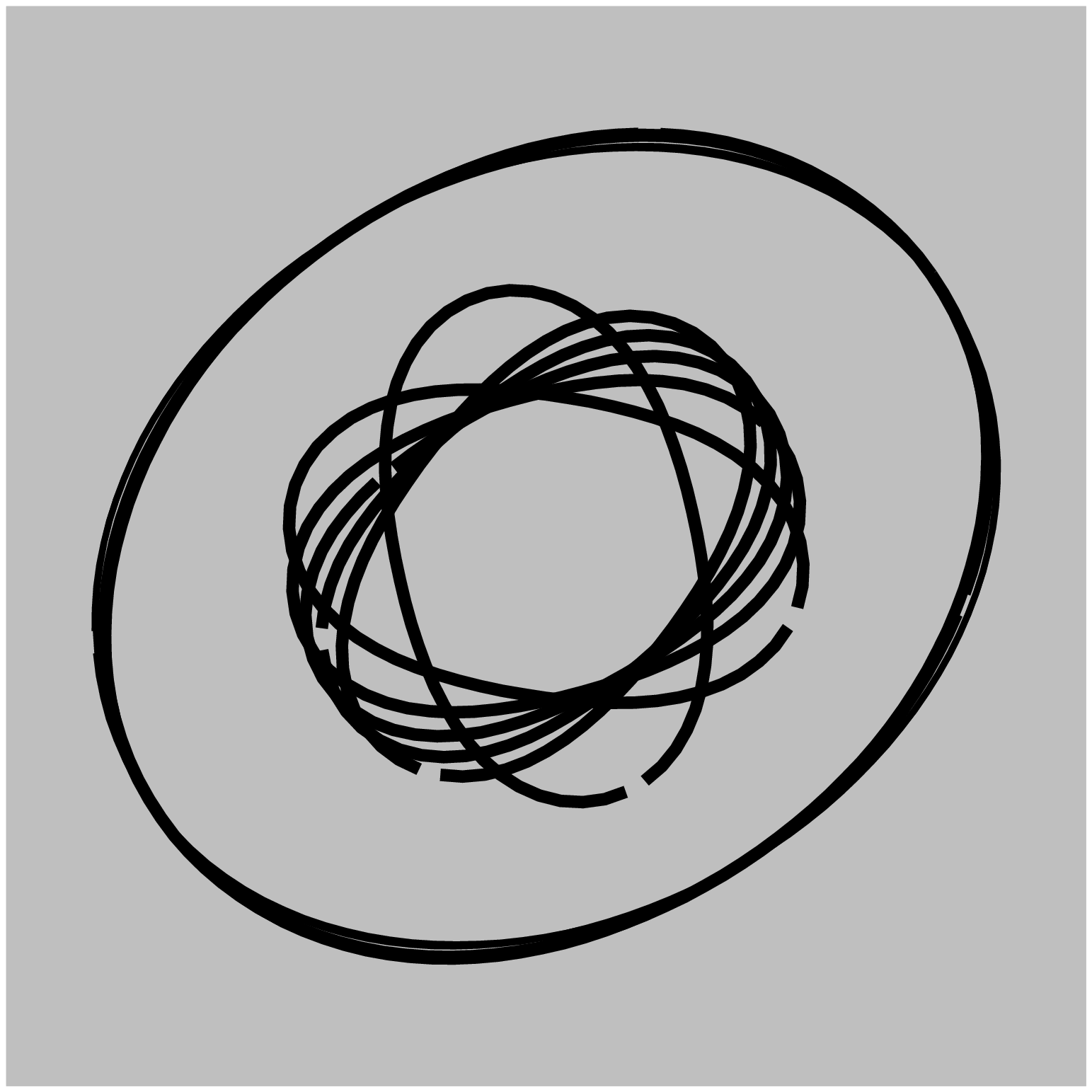}}
\quad
\subfloat[$\Delta_\text{out} = 1.00$]{\label{fig:figure07c}\includegraphics[angle=-90,width=0.14\textwidth]{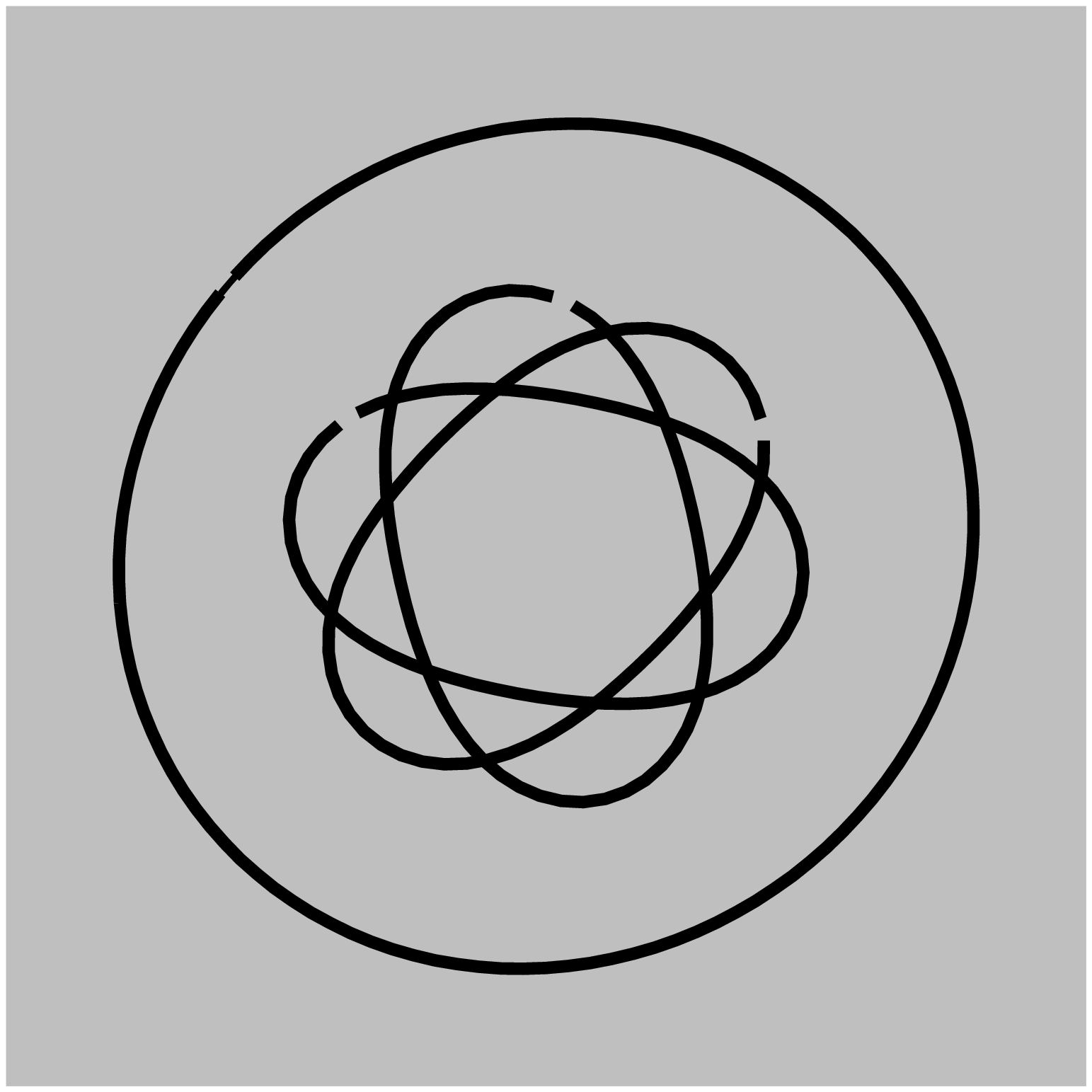}}
\\
\subfloat[$\Gamma / S = 11.56$]{\label{fig:figure07d}\includegraphics[angle=-90,width=0.14\textwidth]{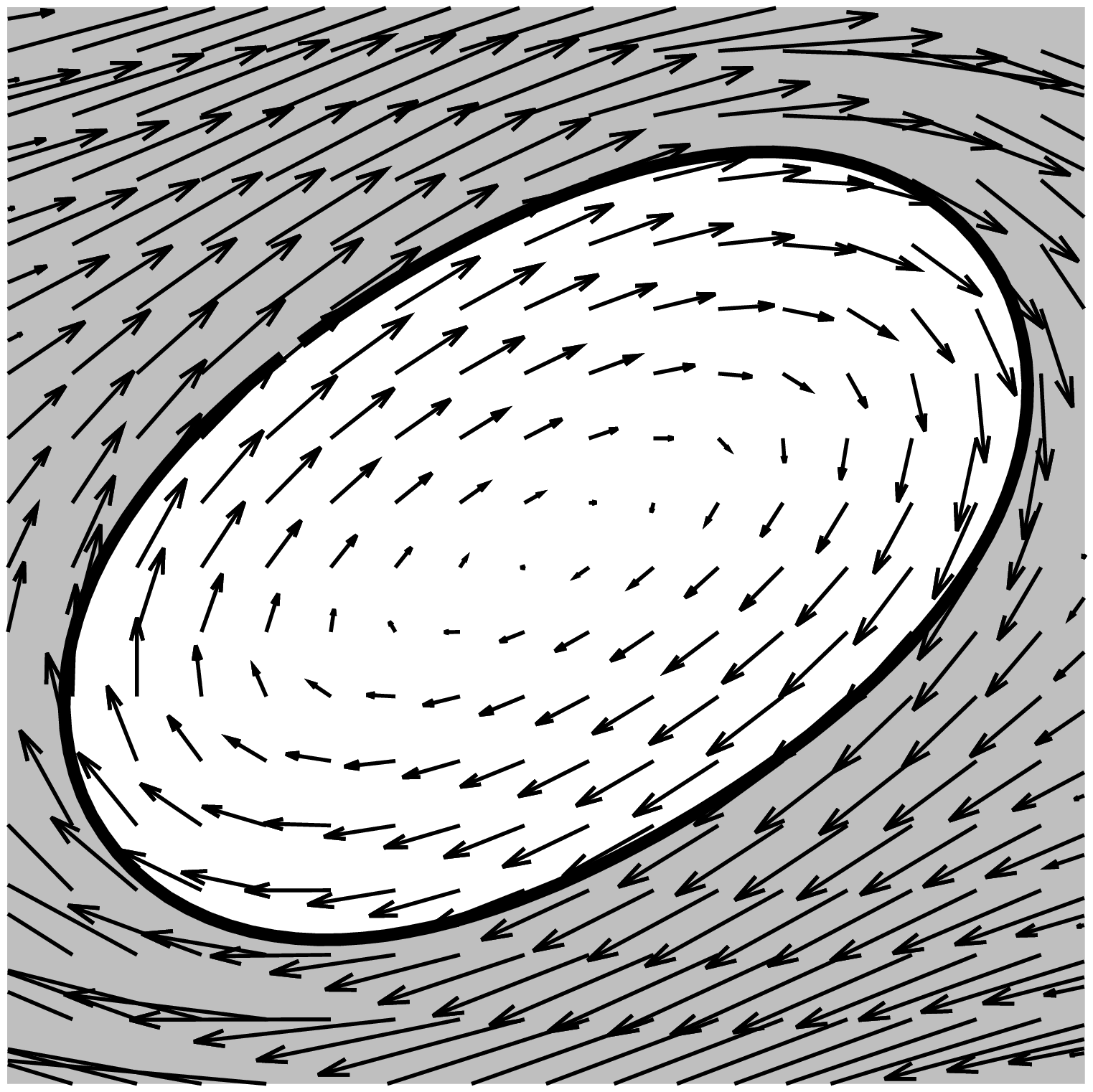}}
\quad
\subfloat[$\Gamma / S = 27.62$]{\label{fig:figure07e}\includegraphics[angle=-90,width=0.14\textwidth]{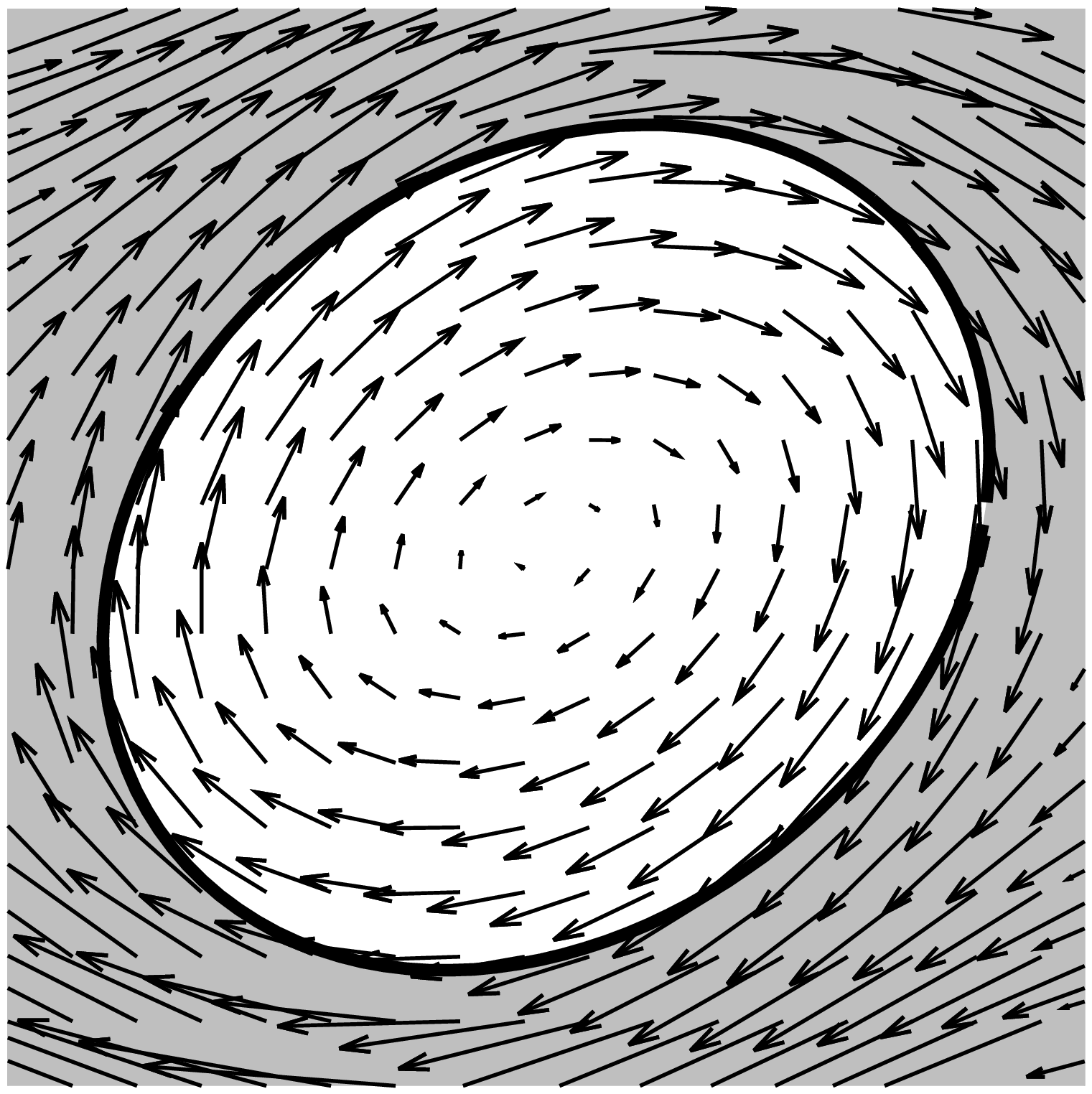}}
\quad
\subfloat[$\Gamma / S = 243.90$]{\label{fig:figure07f}\includegraphics[angle=-90,width=0.14\textwidth]{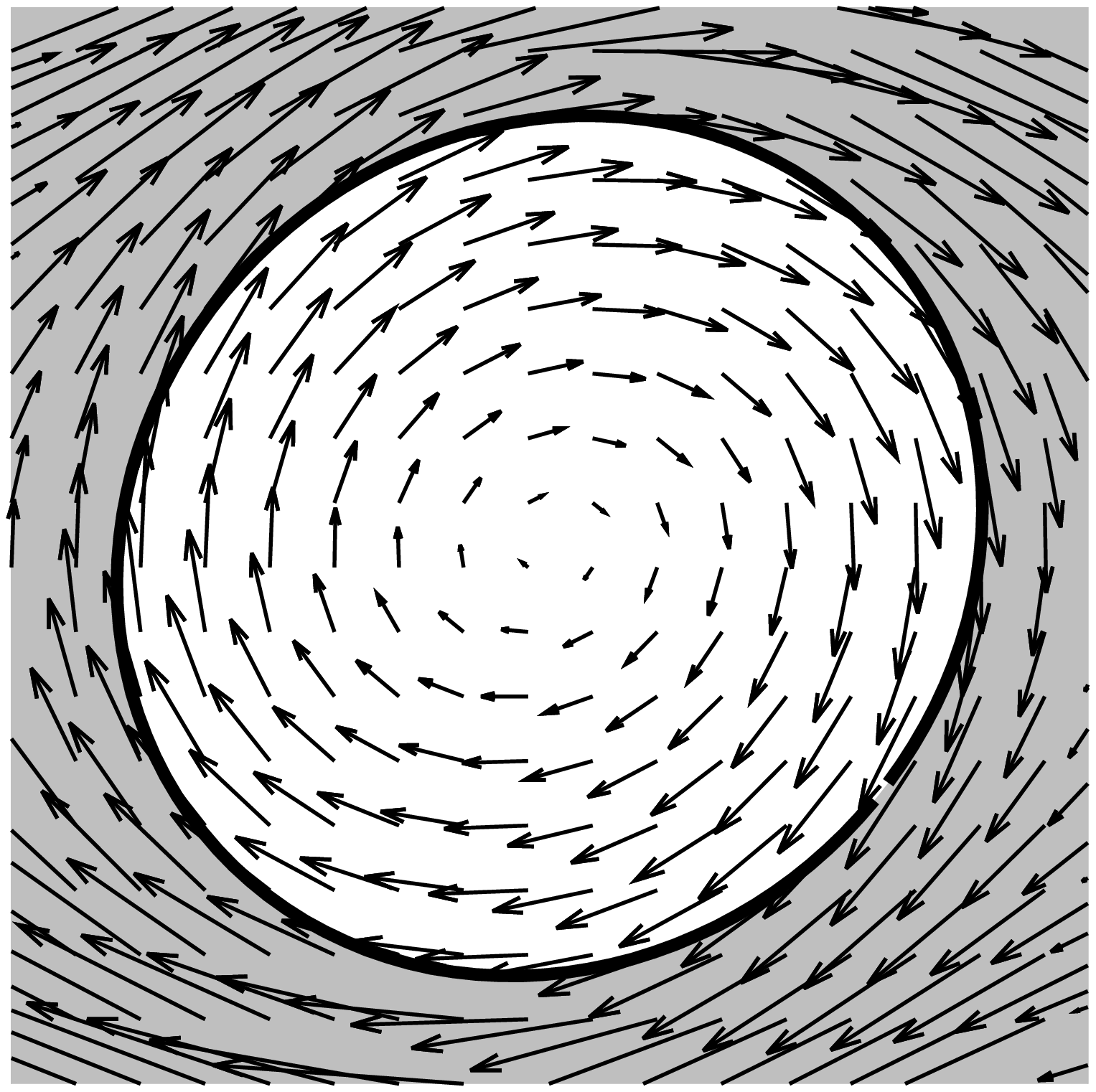}}
\caption{\label{fig:figure07} The dynamical states of the inner vesicle as
a function of the outer vesicle swelling degree $\Delta_\text{out}$ at ${\rm Ca}=0.5$. (a)
Tank-treading motion, (b) tumbling while the
outer vesicle performs a breathing-like motion, and (c) tumbling with a
constant angular velocity. Panels (d)--(f) show the resulting flow field
(in absence of an inner vesicle). It evolves from mixed flow to pure
rotational flow when $\Delta_\text{out} \to 1$. $\Gamma / S$ denotes the ratio of flow
vorticity and shear rate.}
\end{figure}

So far we described how the inner vesicle alters the dynamics of the outer
vesicle and so of the BLV. Next, we examine how the outer vesicle
in turn influences the dynamics of the inner vesicle. This is
controlled by the amount of the fluid between the two membranes. To understand its relevance
we vary its amount by
swelling (adding fluid) or deflating (removing fluid) the outer vesicle.
We vary $\Delta_\text{out}$ between $0.9$ and $1$, keeping all other
parameters unchanged. Consequently, we observe another TT-TB transition,
but this time for the inner vesicle as shown in
Fig.~\ref{fig:figure07}(a)--(c).  For $\Delta _\text{out}= 0.9$, both vesicles
tank-tread (Fig.~\ref{fig:figure07}(a)). Above a critical value of
$\Delta_\text{out}=0.98$, the inner vesicle starts tumbling while
the outer experiences a breathing-like motion
(Fig.~\ref{fig:figure07}(b)). 
For $\Delta_\text{out}=1$ (Fig.~\ref{fig:figure07}(c)), the BLV behaves
exactly as a solid body. The two vesicles rotate with the same angular
velocity and also the fluids, between the membranes and within the inner
membrane, behave like a solid medium. The dynamical transition observed
here for the inner vesicle is expected to modify the apparent viscosity
within the BLV. A similar link between the rheology and
the micro-dynamics has been observed for red blood cells \cite{Forsyth2011}. Although the
BLV dynamics for the three cases (Fig.~\ref{fig:figure07}(a)--(c)) is
apparently similar for an outside observer, their rheological properties
may differ due to the dynamical state (TT or TB) of the inner vesicle.
The dynamical transition induced by varying $\Delta_\text{out}$ alone can be
explained by the theory of Lebedev \textit{et al.} \cite{Lebedev2007} who predicted that even a non-viscous vesicle ($\Lambda = 1$) can undergo a TT-TB
transition by increasing the rotational component of the external
imposed flow.  This was later confirmed experimentally by Deschamps
\textit{et al.} \cite{Deschamps2009}.  In our case, the inner vesicle is
subjected to the flow induced by the tank-treading motion of the outer
membrane.
Fig.~\ref{fig:figure07}(d)--(f) depict the generated undisturbed flow (in the absence of the inner vesicle) for each $\Delta_\text{out}$.
A pure rotational flow is obtained for $\Delta_\text{out} = 1$.
For $\Delta_\text{out} = 0.9$ and $\Delta_\text{out} = 0.98$, a mixed flow is generated, i.e., a combination of pure shear and pure rotational flows.
We quantify the relative importance of the rotational and the elongational components using the quantity $\Gamma / S$, where $\Gamma$ is the vorticity and $S$ is the shear magnitude ($\Gamma / S = 1$ for pure shear and $\Gamma / S \rightarrow \infty$ for pure rotational flow).
The computed values $\Gamma / S$ for each $\Delta_\text{out}$ are shown in Fig.~\ref{fig:figure07}(d)--(f).
The TT-TB transition is induced by increasing $\Gamma / S$ beyond the critical value of $27.62$ (corresponding to the critical value $\Delta _{\rm out}=0.98$).
Using Lebedev's parameters, $\overline{S} = 7 \pi \Canum / \sqrt{3} \delta$ and $\overline{\Lambda} = 4(1 + 23 \Lambda / 32) \sqrt{\delta} / \sqrt{30 \pi}$ ($\delta$ is the excess perimeter in 2D or excess area in 3D), the transition occurs in our case at $\overline{S} _\text{cr} = 0.42$ and $\overline{\Lambda} _\text{cr}= 11.41$.
This $\overline{\Lambda} _\text{cr}$ is larger than expected when compared to the case of a viscous vesicle under shear flow ($\overline{\Lambda}_\text{cr} = 1.38$) \cite{Kaoui2012} or a non-viscous vesicle in general flow ($\overline{\Lambda}_\text{cr} = 1.2$) \cite{Deschamps2009}.
At low deformation ($\overline{S} \rightarrow 0$), $\overline{\Lambda}_\text{cr}$ is independent of $\overline{S}$ anyway \cite{Lebedev2007,Deschamps2009}.
The discrepancy in $\overline{\Lambda} _\text{cr} \varpropto \Lambda$ can be explained by taking into account that Lebedev's theory has been formulated for unbounded flows while, in our case, the inner vesicle is strongly confined by the outer vesicle membrane.
Confinement is found to shift the critical point of the TT-TB transition to larger values of $\Lambda$ \cite{Kaoui2012}.
Thus, Lebedev's theory can explain only qualitatively the TT-TB transition observed for the inner vesicle in Fig.~\ref{fig:figure07}.

\section{Conclusions} 
We showed that a non-viscous vesicle
exhibits rich complex dynamics
when it encapsulates another non-viscous vesicle. Increasing the size of the
inner vesicle triggers a dynamical transition from TT to TB or the newly found
\textit{undulating} motion. The BLV internal medium
displays non-Newtonian behavior with a time-dependent apparent
viscosity during unsteady motion: the same BLV behaves like a solid or a fluid
depending on its orientation and the dynamical state of its inner vesicle. Our
results suggest that a leukocyte cannot be simply mimicked with a solid
spherical particle or with an inclusion-free vesicle enclosing a homogenous
Newtonian fluid. The presence of an internal structure dictates its dynamical
and rheological response to imposed flow. Our results suggest consequences on
the margination and adhesion of leukocytes in the microcirculation, and its
physiological and pathological implications. Along this study we approximated the two fluids, mimicking the cytoplasm and the nucleus, to be simple Newtonian with identical viscosity. This is a simplistic picture when compared to the actual complex nature of the internal structure of a leukocyte (presence of actine, microtubules, filaments). By considering the cytoplasm as a visco-elastic medium or a more viscous nucleus, we expect this to lead to different dynamical behaviors which are not captured by the present model because this would affect the way the outer and the inner membrane interact hydrodynamically. This will be the subject of future research. In the present model, the considered membranes have only bending properties, while in real 3D systems the membrane shear elasticity comes into play. This may lead to the appearance of new dynamical states, wrinkling of the membrane, or the formation of more than four lobes.

\begin{acknowledgments}
We thank the NWO/STW for financial support (VIDI grant 10787
of J. Harting) and the anonymous referees for valuable comments.
\end{acknowledgments}

%
%

%
\end{document}